%% file: 0_Main.tex
\documentclass[preprint,12pt]{elsarticle}

\usepackage{hyperref}
\usepackage{amssymb}
\usepackage{amsmath}
\usepackage{svg}
\usepackage{lineno}
\usepackage{booktabs}
\usepackage{algorithm}
\usepackage{algpseudocode}
\usepackage{comment}
\usepackage[top=2.5cm, bottom=2cm, left=2cm, right=2cm]{geometry}
\usepackage{subcaption}
\usepackage[colorinlistoftodos]{todonotes}

\DeclareMathOperator*{\argmin}{arg\,min}
\DeclareMathOperator{\Rr}{\mathbb{R}}
\newcommand{\Rmat}[2]{\Rr^{#1 \times #2}}
\newcommand{\Rvec}[1]{\Rr^{#1}}

\def \dd {\text{d}} 

\journal{Brain Multiphysics}

\begin{document}
\begin{frontmatter}

\title{ Geometry Reduced Order Modeling (GROM) with application to modeling of glymphatic function}

\author[SCAN,Math]{Andreas Solheim \corref{cor1}}
\ead{solheim@simula.no}
\cortext[cor1]{Corresponding author:}

\author[Jebsen,Radiology,Clinical,Geriatric]{Geir Ringstad}
\author[Jebsen,Clinical,Neurosurgery]{Per Kristian Eide}
\author[SCAN,Math,Jebsen]{Kent-Andre Mardal}

\affiliation[SCAN]{organization={Department of Numerical Analysis and Scientific Computing (SCAN), Simula Research Laboratories},
             addressline={Kristian Augusts gate 23},
             city={Oslo},
             postcode={0164},
             country={Norway}}
\affiliation[Math]{
organization={Department of Mathematics, University of Oslo},
            addressline={Moltke Moes vei 35},
             city={Oslo},
             postcode={0851},
             country={Norway}}

\affiliation[Jebsen]{
organization={K.G. Jebsen Centre for Brain Fluid Research, University of Oslo},
addressline={Pb 1072 Blindern},
 city={Oslo},
 postcode={0316},
 country={Norway}}

\affiliation[Radiology]{
organization ={Department of Radiology, Oslo University Hospital},
addressline={Pb 4950 Nydalen},
city={Oslo},
postcode={0424},
country={Norway}
}

\affiliation[Clinical]{
organization ={Institute of Clinical Medicine, University of Oslo},
addressline={Klaus Torgårds vei 3},
 city={Oslo},
 postcode={0372},
 country={Norway}}

\affiliation[Geriatric]{
organization ={Department of Geriatric Medicine, Sørlandet Hospital Trust},
addressline={Pb 416 Lundsiden},
 city={Arendal},
 postcode={4604},
 country={Norway}}

\affiliation[Neurosurgery]{
organization ={Department of Neurosurgery,
Oslo University Hospital-Rikshospitalet},
addressline={Sognsvannsveien 20},
 city={Oslo},
 postcode={0372},
 country={Norway}}

\begin{abstract}
Computational modeling of the brain has become a key part of understanding how the brain clears metabolic waste, but patient-specific modeling on a significant scale is still out of reach with current methods. We introduce a novel approach for leveraging model order reduction techniques in computational models of brain geometries to alleviate computational costs involved in numerical simulations. Using image registration methods based on magnetic resonance imaging, we compute inter-brain mappings which allow previously computed solutions on other geometries to be mapped on to a new geometry. We investigate this approach on two example problems typical of modeling of glymphatic function, applied to a dataset of $101$  MRI of human patients. We discuss the applicability of the method when applied to a patient with no known neurological disease, as well as a patient diagnosed with idiopathic Normal Pressure Hydrocephalus displaying significantly enlarged ventricles

\textit{Statement of significance}: In many fields, model order reduction is a key technique in enabling high-throughput numerical simulations, but remains largely unexploited for biomedical modeling of the brain. In this work, we introduce a novel technique for building reduced representations integrating simulations performed on other brain geometries derived from MRI. Using this technique, we may leverage a dataset of previous solutions to accelerate simulations on new geometries, making patient-specific modeling more feasible.

\end{abstract}


\end{frontmatter}


\input{1_Introduction}
\input{2_Methods}
\input{3_0_Results}
\input{4_Discussion}
\input{5_Conclusion}

\input{7_Supplementary_Declarations}

\bibliographystyle{elsarticle-num-names} 
\bibliography{ref}

\end{document}

%% file: 1_Introduction.tex
\section{Introduction}

The fluid-structure interaction processes of the human brain has in recent years become an important research topic due to its relation to solute transport and brain clearance. In particular, the glymphatic system is a brain-wide perivascular transport route that provides for CSF-mediated solute and fluid transport within the brain. Adding to the discovery of the glymphatic system~\citep{Iliff2012}, was the characterization of meningeal lymphatic vessels capable of draining CSF~\citep{Louveau2015, Aspelund2015}. The CSF transport is crucial for brain health, and in particular its clearance of metabolic waste which seems to be facilitated during sleep~\cite{xie2013sleep, bojarskaite2023sleep, eide2021sleep}. Reduced clearance is linked to the accumulation of toxic proteins such as amyloid-$\beta$ and tau, which is the hallmark of Alzheimer's disease~\citep{Hardy2002, Spillantini2013}, as well as Lewy-bodies and $\alpha$-synuclein which has been tied to Parkinson's disease especially~\citep{Goedert2001}. 

Multiple computational models have been proposed for assessing glymphatic clearance
\citep{corti2024structure, dreyer2024modeling, fumagalli2024polytopal, Hornkjol2022, johnson2023image, Valnes2020-jh, vardakis2020exploring, vinje2023human}. Common for these studies is that only few subjects are studied as multi-physics processes of subject specific models are currently a formidable computational challenge. For instance, in~\citet{Hornkjol2022} around 30 000 CPU hours were used for a single subject considering the solute transfer and fluid flow of CSF in order to investigate the influence of CSF mediated solute transport on glymphatic function. Similarly, in~\citet{vinje2023human} approximately 64 000 CPU hours were used to assess the influence of sleep and sleep deprivation in 24 subjects. As such, there is a need for computational approaches that enable larger cohorts of subjects. 

Model order reduction (MOR) is a technique which aims to alleviate the computational cost of performing simulations by leveraging previous simulations~\citep{chinesta2016, hesthaven2016certified}, often called snapshots. With our pipeline~\cite{mri2fem_1}, magnetic resonance images (MRIs) are segmented via FreeSurfer~\cite{dale1999}, meshed in SVMTK and exported to FEniCS~\cite{logg2012automated} format with proper subdomain markers and adjustable resolution. A main issue, however, is the fact that the subject specific meshes will differ in terms of geometry, as well as the number of degrees of freedom. As such, a main challenge to enable MOR is to construct mappings that are flexible and robust with respect to inconsistencies between the meshes. 

Due to the direct relationship between the underlying MRI and the computational mesh, we propose an approach to this challenge which relies on image-based registration methods. Image registration is an important tool in radiology, as it allows for more direct inter- and intra-subject comparisons of imaging data. Accordingly, many approaches have been proposed for brain imaging specifically~\citep{fischer2003flirt, andersson2008fnirt, ASHBURNER2007, ASHBURNER2011, Hoffmann2022, Postelnicu2009, AVANTS2008, Modat2014-er}, see~\citet{Klein2009} and~\citet{Klein2010} for a detailed comparison of some of these methods. In this work, we apply the image registration method introduced in~\citep{AVANTS2008}, available from the Advanced Normalization Tools (ANTs) image toolbox, to $T_1$-weighted brain MRI. We thus compute image-to-image mappings to act as a vehicle for inter-subject brain transformations which can be applied to meshes derived from MRI. This means that the inter-geometry transformation is completely independent of the construction of the mesh representing the brain geometry, but rather entirely tied to the underlying MRI.

Our approach is tested and implemented on a dataset of $101$ $T_1$-weighted brain MRI of human patients. All patients in the dataset were imaged while under investigation for suspected CSF disorders, although not all subjects were found to have a defined CSF disorder or neurological disease. The main pathology present in this dataset is patients diagnosed with idiopathic Normal Pressure Hydrocephalus (iNPH), which is a combined CSF and neurodegenerative disease with overlap toward Alzheimer's disease~\cite{Eide2025}. In patients with iNPH, disturbances in CSF flow is associated with enlarged ventricles~\cite{Malm2006} and thus fundamentally changes the geometry of the brain. As such, we compare the performance of the method on an iNPH patient with another patient from a group of reference (REF) patients. After thorough diagnostic assessment, the REF individuals had no identified CSF disturbance or other neurological disease, and were therefore considered as close to healthy. The method is illustrated on two example problems typical of numerical modeling of glymphatic function.

%% file: 2_Methods.tex
\section{Geometric model order reduction} \label{sec:methods}
\begin{figure*}[t]
    \centering
    \includegraphics[width=\linewidth]{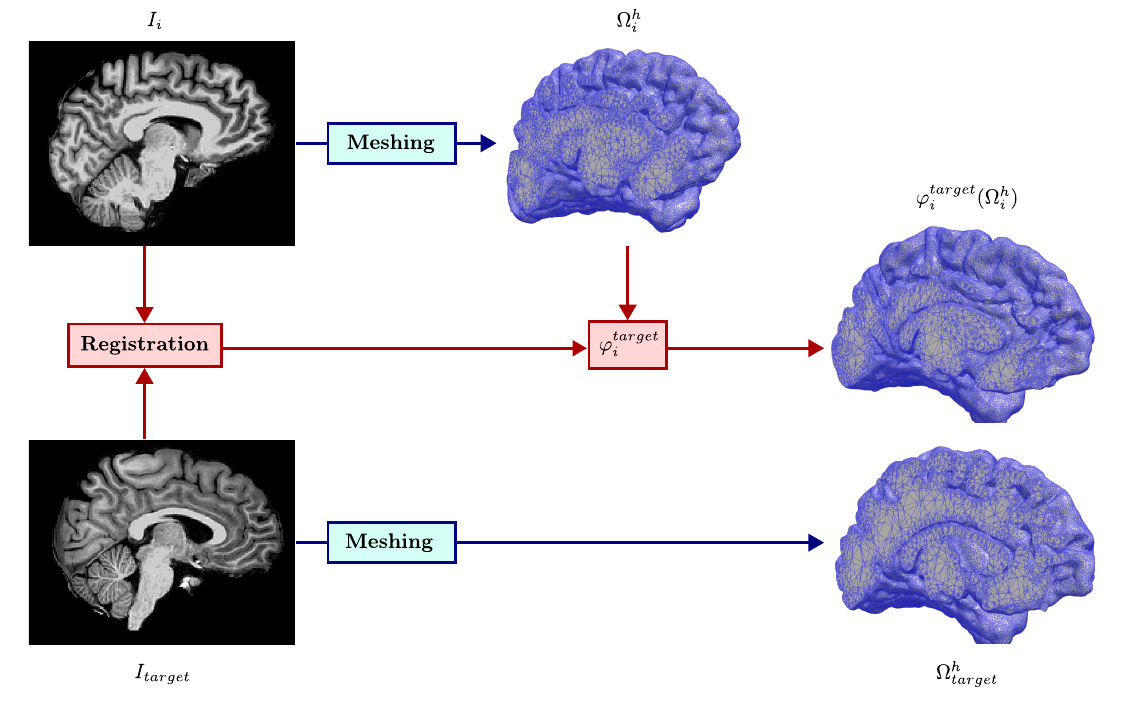}
    \caption{Image registration applied to meshes: Registration is performed by matching the MRI $I_i$ (top left) to $I_{target}$ (bottom left), thus obtaining $\varphi_i^{target}$. Using meshing tools we derive volumetric meshes from MRI, and apply $\varphi_i^{target}$ to the mesh $\Omega_i^h$ (top) to obtain the deformed mesh $\varphi_i^{target}(\Omega_i^h)$ (middle right), which approximates the target mesh $\Omega_{target}^h$ (bottom right).}
    \label{fig:diagram}
\end{figure*}
\subsection{Image registration} \label{sec:IR}

The goal of image registration is to compute a mapping from one image to another, by matching $T_1$-weighted MRI signal intensities $I_i$ to a target image $I_{target}$. We consider voxelized images on a $256\times 256\times 256$ grid, with a uniform $1$~mm resolution. Through optimization, a mapping $\varphi_i^{target}$ from image $I_i$ to $I_{target}$ is computed such that $I_i \circ \varphi_i^{target} \approx I_{target}$. The concept of applying the deformation $\varphi_i^{target}$ obtained by image registration to a finite element mesh rests on the fundamental correspondence between the MRI of a brain $I_i$, represented by an image, and any associated mesh $\Omega_i^h \subset \Rvec{3}$, represented by a set of points. Brain surface meshes are obtained through an iterative process of approximating surface features from a segmentation of the gray cortical matter and ventricles, and accordingly the mesh of each brain is unique. As these meshes describe complex surfaces with significant curvature, they are highly unstructured, and we do not expect the points of the mesh to correspond exactly with the voxels of the image. Depending on the mesh parameter $h$, the finite element mesh may also be locally far more dense or far less dense than the image. The critical advantage of using images for registration rather than meshes is that every image is defined on the same grid, while mesh nodes will not necessarily correspond. 

Consequently, when $\varphi_i^{target}$ is applied to $\Omega_i^h$, we do not recover $\Omega_{target}^h$, but rather an approximation of the target mesh. We refer to Figure~(\ref{fig:diagram}), which illustrates how the mesh $i$ is mapped to the target mesh. We notice that the deformed mesh $\varphi_i^{target}(\Omega_i^h)$ outlines the same geometry as $\Omega_{target}^h$, however the two meshes are distinct as the mesh nodes do not correspond.

\subsection{Model order reduction} \label{sec:MOR}

We apply an offline-online approach to model order reduction based on the precomputing of a set of high-fidelity snapshot solutions~\citep{Sirovich_1991}. We consider a situation where many images $I_i$ and associated meshes $\Omega_i^h$ are available, allowing us to precompute solutions on these geometries during the offline phase. For any given problem and associated parameters, this process only needs to be done once. During the online phase, we consider a situation in which we want to solve the problem of interest on a new, unseen target brain geometry $\Omega_{target}^h$. Based on the image representation of this brain geometry, we compute the mapping of each brain in the dataset to the target geometry. Using this dataset of previous solutions, we may thus compile a set of snapshot solutions on the geometry of interest, and compute a low-dimensional basis representation using the Proper Orthogonal Decomposition (POD)~\citep{volkwein2011model}. During the online phase, the same underlying problem is solved in the reduced basis representation, significantly reducing the computational cost of the forward problem.   

\subsubsection{Precomputing snapshots} \label{sec:snapshots}

Consider a general variational problem, with a bilinear form $a(\cdot, \cdot)$, defined on $V_i \subseteq H(\Omega_i)$, where $H$ is a Hilbert space for a geometry $\Omega_i \subset \Rvec{3}$. Assume that $a(\cdot, \cdot)$ is continuous and coercive on $V_i$. Let $u_i \in V_i$ be such that:
\begin{equation} \label{eq:variational}
    a_i(u_i,v_i) = f_i(v_i) \quad \forall v_i \in V_i
\end{equation}
Where $f_{i}(\cdot)$ is the linear form of the PDE. This problem can be solved numerically by introducing a discretization of non-degenerate triangulations $\Omega_i^h$ of $\Omega_i$, which we refer to as a mesh. 
Different meshes will have different numbers of degrees of freedom, and
a direct mapping between meshes will, in general, be degenerate. Consequently, the function spaces $V_i$ also have a different number of degrees of freedom and cannot be mapped directly to each other. 
From each mesh, we can build a conforming subspace $V_i^h \subseteq V_i$ and solve the discrete variational problem of finding $u_i^h \in V_i^h$ such that:
\begin{equation} \label{eq:numerical} 
    a_i(u_i^h,v_i) = f_i(v_i) \quad \forall v_i \in V_i^h
\end{equation}
The discrete problem can then be solved for each geometry $\Omega_i^h$ and generating a dataset of solutions on $M$ available geometries:
\begin{equation*}
    \{u_i^h \in V_i(\Omega_i^h)\}_{i=1}^M
\end{equation*}
We aim to map these solutions to the target geometry $\Omega_{target}^h$. 
For each image-mesh pair in the dataset, we thus compute the mapping $\varphi_i^{target}$, based on the MRI $I_i$ and $I_{target}$, and subsequently apply the transformation to each precomputed solution $u_i$.

This procedure yields a set of solutions on the deformed geometries $\varphi_i^{target}(\Omega_i^h)  = \Tilde{\Omega}_i^h$ which approximate the target geometry $\Omega_{target}^h$:
\begin{equation*}
    \{\Tilde{u}_i^h = u_i^h \circ  \varphi_i^{target} \}
\end{equation*}
A key observation is that the deformed solutions $\Tilde{u}_i^h$ will not be elements of $V_i(\Tilde{\Omega}_i^h)$. The change of variables means that Eq.~(\ref{eq:numerical}) will generally not be satisfied on $\Tilde{\Omega}_i^h$. In order for the variational form to be satisfied on this geometry, we need to modify the differential operators to be consistent with $\varphi_i^{target}$, i.e set: 
\begin{equation} \label{eq:geom_sampling}
    \hat{\nabla}_i = \mathcal{J}_{\varphi_i^{target}} \nabla_i \quad \text{and} \quad \text{d} \Tilde{\Omega}_i^h = |\mathcal{J}_{\varphi_i^{target}}| \text{d} \Omega_i^h
\end{equation}
Where $\mathcal{J}_{\varphi_i^{target}}$ denotes the Jacobian of $\varphi_i^{target}$ in each point $\mathbf{x} \in \Omega_i^h$. By defining modified bilinear and linear forms $\hat{a}_i(\cdot, \cdot)$ and $\hat{f}_i$ using these differential operators, as well as a modified function space $\hat{V}_i(\Tilde{\Omega}_i^h)$ each deformed solution $\Tilde{u}_i^h$ will satisfy:
\begin{equation} \label{eq:deformed_variational}
    \hat{a}_i(\Tilde{u}_i^h,\Tilde{v}_i) = \hat{f}_i(\Tilde{v}_i) \quad \forall \Tilde{v}_i \in \hat{V}_i(\Tilde{\Omega}_i^h)
\end{equation}
The act of sampling solutions from various geometries, is therefore essentially equivalent to sampling solutions with modified differential operators.

However, sampling solutions from Eq.~(\ref{eq:deformed_variational}) is itself not enough. As seen in Figure~(\ref{fig:diagram}), while the mesh $\Tilde{\Omega}_i^h$ is a reasonable approximation of $\Omega_{target}$, the cells and nodes of the mesh will still be distinct from $\Omega_{target}^h$, which is where $u_{target}^h$ is defined. We must therefore additionally extend each solution $\Tilde{u}_i^h$ from $\Tilde{\Omega}_i^h$ to $\Omega_{target}^h$. For nodes where the two meshes overlap, this can be directly done through interpolation of the finite element basis. At the boundary, some nodes in $\Tilde{\Omega}_i^h$ may not lie within a cell of $\Omega_{target}$ due to the registration not being exact. In such cases, we perform extrapolation by gaussian weighting of the closest neighboring cell values. We use the notation $\hat{u}_i^h$ to denote the extension of the deformed solution $\Tilde{u}_i^h$ from $\Tilde{\Omega}_i^h$ to $\Omega_{target}^h$. 

\subsubsection{Constructing the reduced basis} \label{sec:RB}

Once the dataset of snapshots on the target geometry $\Omega_{target}^h$ has been computed, we aim to use these solutions to solve the underlying problem in a reduced space. In particular, we aim to replace $V_{target}(\Omega_{target}^h)$ by the space spanned by these sample solutions $\hat{V}_{M} = \text{span}\{\hat{u}_1 \dots \hat{u}_M\}$ to solve Eq.~(\ref{eq:numerical}). The purpose of the POD is to construct an optimal orthonormal basis for this space. 

We compute this basis by the following approach: Each snapshot has $D$ degrees of freedom, given by the mesh $\Omega_{target}^h$ and the order of the finite element space of choice. When each solution contains $n$ fields, with $D$ degrees of freedom, we treat each field as independent and consider the solution to contain $\Tilde{D}=n\cdot D$ degrees of freedom. We express each solution as a vector with $\Tilde{D}$ components and stack them column-wise to construct a snapshot matrix $\mathbf{X} \in \Rmat{\Tilde{D}}{M}$:
\begin{equation*}
    \mathbf{X} = 
    \begin{pmatrix}
        |&|&|&| \\
        \hat{u}_1  & \hat{u}_2 &  \cdots & \hat{u}_M \\
        |&|&|&| \\
    \end{pmatrix}
\end{equation*}
Computing a low-dimensional representation of this data using the POD corresponds to finding the $d$-rank matrix $\mathbf{U}$, such that:
\begin{equation} \label{eq:low-rank}
    \mathbf{U} = \argmin_{\mathbf{U} \in \Rmat{\Tilde{D}}{d}} || \mathbf{X} - \mathbf{U} \mathbf{U}^\top \mathbf{X} ||_F \quad \mathbf{U}^\top \mathbf{U} = \mathbf{I}
\end{equation}
Where $\mathbf{I} \in \Rmat{d}{d}$ denotes the $d \times d$ identity matrix. By computing such a matrix $\mathbf{U}$ with rank $d << \Tilde{D}$, we can obtain a very low-dimensional orthogonal basis of the space spanned by $\mathbf{X}$. From the Eckart-Young theorem~\citep{Eckart1936}, the optimal solution to~(\ref{eq:low-rank}) is to compute the SVD of the matrix:
\begin{equation*} \label{eq:svd_error}
    \mathbf{X} = \hat{\mathbf{U}} \mathbf{\Sigma} \hat{\mathbf{V}}^\top
\end{equation*}
The optimal $d$-rank basis representation of $\mathbf{X}$ can be found by extracting the first $d$ columns of the left-side singular matrix, setting $\mathbf{U} = \hat{\mathbf{U}}_{1:d}$. The matrix $\mathbf{\Sigma} \in \Rmat{S}{M}$, $S=\min(\Tilde{D},M)$ is the singular value matrix, which consists entirely of $0$'s, except for the diagonal of the upper $S\times S$ sub-matrix which contains the singular values $\{\sigma_i\}_{i=1}^S$. The singular values are key to understanding the POD as they measure the error of the $d$-rank approximation of $\mathbf{X}$~\citep{Eckart1936}:
\begin{equation*}
 || \mathbf{X} - \mathbf{U} \mathbf{U}^\top \mathbf{X} ||_F = \sum_{i=d+1}^S \sigma_i
\end{equation*}

Having obtained $\mathbf{U}$, we subsequently seek to solve~(\ref{eq:numerical}) in the low-dimensional space spanned by this basis. This involves finding $u_{target}^{rb} \in V_{target}^{rb}$, where $V_{target}^{rb}$ is defined as:
\begin{equation*}
    V_{target}^{rb} = \text{span}\{\mathbf{U}_1 \dots \mathbf{U}_d\}
\end{equation*}
We then subsequently solve the reduced variational problem:
\begin{equation} \label{eq:reduced}
    a_{target}(u_{target}^{rb},v_{target}) = f_{target}(v_{target}) \quad \forall v_{target} \in V_{target}^{rb}
\end{equation}
In practice, we can obtain~(\ref{eq:reduced}) from~(\ref{eq:numerical}) by projection:
\begin{align*}
    A_{target}^{rb} &= \mathbf{U}^\top A_{target}\mathbf{U} &\quad \left(A_{target}^{rb}\right)_{i,j} &= a_{target}(\mathbf{U}_i,\mathbf{U}_j) \\
    F_{target}^{rb} &= \mathbf{U}^\top F_{target} &\quad \left(F_{target}^{rb}\right)_i &= f_{target}(\mathbf{U}_i)
\end{align*}
The reduced basis solution $u_{target}^{rb}$ can now be obtained by solving $A_{target}^{rb}u_{target}^{rb} = F_{target}$. The high-dimensional representation can then be recovered by inverse projection $u_{target} = \mathbf{U}u_{target}^{rb}$.

\section{Models and data}

\subsection{Models} \label{sec:Models}

\subsubsection{Two-compartment model of tracer distribution}

As a first example, we implement the model proposed in~\citet{Riseth2024}, at steady state. This is a diffusion-dispersion continuum model, simulating exchange of MRI tracer concentration between the extracellular space (ECS) and perivascular space (PVS) surrounding blood vessels in the brain, as well as some clearance into blood from perivascular spaces. The model assumes that the tracer concentration in blood and advective transport is negligible, and that tracer molecules may only disperse from the PVS into blood, but not directly from the ECS. Concentration at the fluid filled subarachnoid spaces (SAS) surrounding the brain are modeled as Robin boundary conditions. We implement the model at steady-state, estimating the peak in concentration, normally seen $15-20$h after injection. The model is formulated as the following set of differential equations: 
\begin{equation}
\begin{aligned} \label{eq:twocomp}
    -\nabla \cdot (n_e D_e \nabla c_e) &= \pi_{ep}(c_p - c_e) & \text{ in }\Omega \\
    -\nabla \cdot (n_p D_p \nabla c_p) &= -\pi_{ep}(c_p - c_e) - \pi_{pb}c_p & \text{ in }\Omega \\
    - n_e D_e \nabla c_e \cdot \mathbf{n} &= k_e (c_e - c_{SAS}) & \text{ on }\partial \Omega \\
    - n_p D_p \nabla c_p \cdot \mathbf{n} &= k_p (c_p - c_{SAS}) & \text{ on }\partial \Omega
\end{aligned}
\end{equation}
Where $c_e$ and $c_p$ denote the concentration in the extracellular and perivascular spaces, respectively. The concentration $c_{SAS}$ is the concentration in the fluid-filled spaces at the boundary of the domain. A complete list of model parameters and values are listed in Table~(\ref{tab:twocomp}). We consider an idealized case of tracer concentration in the SAS, which we estimate as~\cite{Riseth2024}:
\begin{equation*}
    c_{SAS}^{\alpha} = a_{\alpha}\phi^{-1} \left(-e^{-t/\tau_1} + e^{-t/\tau_2}\right)
\end{equation*}
Where $\alpha\in\{$pial, ventricle\}, with $a_{\text{pial}}=0.52$mm$^2$/s and $a_{\text{ventricle}}=0.2$mm$^2$/s, $\phi=0.2$ is the volume fraction, $\tau_1=4.43 \cdot 10^4$s and $\tau_2=8.5 \cdot 10^4$s. We set $t=16.7$h, which is the point where the estimated SAS concentration reaches its peak.  

\begin{table}
\centering
\begin{tabular}{@{}cllr@{}}
\toprule
Quantity   & Description                    & \multicolumn{1}{l}{Value} & \multicolumn{1}{l}{Unit} \\ \midrule
$n_e$      & ECS volume fraction            & $0.2$                     & -                        \\
$n_p$      & PVS volume fraction            & $0.2 \cdot 10^{-1}$       & -                        \\
$D_e$      & ECS diffusion coefficient      & $1.3 \cdot 10^{-4}$       & mm$^2$/s                 \\
$D_p$      & PVS diffusion coefficient      & $3.9 \cdot 10^{-4}$       & mm$^2$/s                 \\
$\pi_{ep}$ & Transfer coefficient ECS-PVS   & $2.9 \cdot 10^{-2}$       & 1/s                      \\
$\pi_{pb}$ & Transfer coefficient PVS-blood & $2.0 \cdot 10^{-8}$       & 1/s                      \\
$k_e$      & ECS surface conductivity       & $1.0 \cdot 10^{-5}$       & mm/s                     \\
$k_p$      & PVS surface conductivity       & $3.7 \cdot 10^{-4}$       & mm/s                     \\ \bottomrule
\end{tabular}
\caption{Model parameters for Eq.~(\ref{eq:twocomp}), using ranges from~\cite{Riseth2024}. We consider a variation of the model with a relatively low rate of clearance to blood.  \label{tab:twocomp}}
\end{table}

\subsubsection{Multi-compartment poro-elastic model (MPET)}

As a second example, we implement a 7-compartment poro-elastic model (MPET) described in~\citet{dreyer2024modeling}, which is a modified version of the model in~\cite{TULLY_VENTIKOS_2011}. A mathematical analysis of a generalized version of the model is provided in~\cite{lee2019mixed}.
This model aims to simulate flow resistance in the brain, and the multi-compartment interactions are relevant for general glymphatic modeling. At steady state, ignoring elastic deformations, the governing equations are given by:
\begin{equation}\label{eq:MPET}
    - \nabla \cdot \frac{\kappa_i}{\mu_i} \nabla p_i = \sum_{i \neq j} \omega_{ij}(p_i - p_j)
\end{equation}
Where $\kappa_i$ and $\mu_i$ model the fluid viscosity and permeability in compartment $i$ and $\omega_{ij}$ models the fluid transfer coefficient between compartment $i$ and $j$, following~\cite{TULLY_VENTIKOS_2011, Guo2019-ss}. We refer to~\citet{dreyer2024modeling} for a detailed explanation of each physical quantity, and considerations on relevant ranges estimated in the literature. In this work, we use the base variant described in~\cite{dreyer2024modeling}, and we summarize the model parameters in Table~(\ref{tab:MPET}). We model the pressure in the capillary ($p_c$), arterial ($p_a$) and venous ($p_v$) blood compartments, as well as the perivascular spaces ($p_{pc},p_{pa},p_{pv}$) associated with each compartment, and the extracellular space ($p_e$). 

In the pre-infusion phase the, boundary conditions associated with Eq.~(\ref{eq:MPET}) are given by:
\begin{equation}
\begin{aligned} \label{eq:MPET_bc}
    \kappa_a \nabla p_a \cdot \mathbf{n} &= Q_{in} \quad  & x \in \partial \Omega_{pial} \\
    \kappa_c \nabla p_c \cdot \mathbf{n} &= -Q_{prod} \quad & x \in \partial \Omega_{ventricle} \\
    \kappa_v \nabla p_v \cdot  \mathbf{n} &= \beta_1 \left ( \frac{p_{DS} + p_{CSF}}{2} - p_v \right) \quad &  x \in \partial \Omega_{pial} \\
    \kappa_{pa} \nabla p_{pa} \cdot \mathbf{n} &= \beta_2 (p_{CSF} - p_{pa}) \quad &  x \in \partial \Omega_{pial} \\ 
    \kappa_{pv} \nabla p_{pv} \cdot  \mathbf{n} &= \beta_3 \left ( \frac{p_{DS} + p_{CSF}}{2} - p_{pv} \right) \quad &  x \in \partial \Omega_{pial} 
\end{aligned}
\end{equation}
Where $Q_{in} = B_{in}/ \int \dd \partial\Omega_{pial}$, $Q_{prod} = 0.33 \text{ml/min}$, $p_{DS}=8.4\text{mmHg}$, $\beta_1=\beta_2=10^{-3}$ and $\beta_3=10^{-7}$. In the remaining compartments, we set $\nabla p_i \cdot \mathbf{n} = 0$ on the pial and ventricular surfaces.

\begin{table}
\centering
\begin{tabular}{@{}llll@{}}
\toprule
\begin{tabular}[c]{@{}l@{}}Transfer\\ coefficient\end{tabular} & Value {[}Pa$^{-1}$s$^{-1}${]} &  Permeability & Value [m$^2$]\\ \midrule
$\omega_{a,c}$   & $1.45 \cdot 10^{-6}$& $\kappa_a$ & $3.63 \cdot 10^{-14}$\\
$\omega_{c,v}$   & $8.75 \cdot 10 ^{-6}$&  $\kappa_c$ & $1.44 \cdot 10^{-15}$\\
$\omega_{c,pc}$   & $8.48 \cdot 10 ^{-10}$&   $\kappa_v$ & $1.13 \cdot 10^{-12}$\\
$\omega_{pa,e}$   & $1.86 \cdot 10 ^{-7}$&   $\kappa_e$ & $2 \cdot 10^{-17}$\\
$\omega_{pv,e}$   & $1.65 \cdot 10 ^{-7}$&  $\kappa_{pa}$ & $3 \cdot 10^{-17}$\\
$\omega_{pa,pc}$   & $10 ^{-6}$&$\kappa_{pc}$ & $1.44 \cdot 10^{-15}$\\
$\omega_{pc,pv}$   & $10 ^{-6}$&$\kappa_{pv}$ & $1.95 \cdot 10^{-14}$\\
$\omega_{pc,e}$   & $10 ^{-10}$&  &   \\ \bottomrule
\end{tabular}
\caption{Model parameters for Eq.~(\ref{eq:MPET}), as listed in the base model in~\citet{dreyer2024modeling}. \label{tab:MPET}
}
\end{table}

\subsection{Data and acquisition} \label{sec:dataset}

In this work, we will apply the aforementioned registration techniques to a dataset of $101$ $T_1$-weighted brain MRI. Parts of this dataset has been reported in previous works on glymphatic solute transport assessed with MRI~\cite{Ringstad2017,Ringstad2018-sv}. Participants in the study were all recruited while under clinical work-up for various CSF disorders, and imaged at the Department of Neurosurgery, University Hospital of Oslo-Rikshospitalet according to the protocol described in~\citet{Ringstad2018-sv}. Study subjects underwent intrathecal gadobutrol injection and were imaged using MRI before injection, in addition to several time-points following injection. In this study, we only consider MRI acquired before intrathecal injection. Data examined in this study was collected in the period $2015-2016$ and approved by the Regional Committee for Medical and Health Research Ethics (REK) of Health Region South-East, Norway (2015/96), the Institutional Review Board of Oslo University Hospital (2015/1868) and the National Medicines Agency (15/04932-7). The study was conducted following the ethical standards of the Declaration of Helsinki of 1975 (revised in 1983). Study participants were included after written and oral informed consent. No additional data was collected for this work.

\begin{table*}[t]
\centering
\begin{tabular}{@{}lccccc@{}}
\toprule
Group & $N$ & Age & Sex (M/F) & Height [cm] & Weight [kg] \\ \midrule
\textbf{iNPH}  & $\mathbf{31}$  & $\mathbf{(71\pm7)}$ $\mathbf{[46-80]}$ & $\mathbf{23/8}$& $\mathbf{(176\pm9)}$& $\mathbf{(83 \pm 17)}$\\ 
\textbf{Non-iNPH}  & $\mathbf{70}$  &$\mathbf{(40\pm 13)}$ $\mathbf{[19-72]}$ & $\mathbf{18/52}$ & $\mathbf{(169\pm 21)}$& $\mathbf{(79\pm20)}$ \\
\quad REF   & $18$  &$(35\pm 11)$ $[22-64]$ & $4/14$ & $(163\pm 41)$& $(75\pm25)$ \\
\quad Cysts (pineal, arachnoid)& $26$ & $(39 \pm 13)$ $[19-68]$ & $6/20$ & $(172\pm 8)$ & $(81 \pm 12)$\\ 
\quad Other & $26$ & $(43 \pm 14)$ $[24-72]$ & $8/18$ & $(171\pm 8)$ & $(79 \pm 22)$\\ 
\midrule
\textbf{Total} & $\mathbf{101}$& $\mathbf{(49 \pm 18)}$ $\mathbf{[19-80]}$ & $\mathbf{41/60}$ & $\mathbf{(171\pm 19)}$ & $\mathbf{(81 \pm 25)}$\\ \bottomrule
\end{tabular}
\caption{Characteristic data for subject groups in this study. Other refers to patients diagnosed with other CSF disorders, such as Spontaneous Intracranial Hypotension (SIH), Idiopathic Intracranial Hypertension (IIH)}
\label{tab:subjects}

\end{table*}

In Table~(\ref{tab:subjects}), we detail the patient groups present in the dataset. The largest single patient group are subjects diagnosed with iNPH, which is a neurodegenerative disease with symptoms like urinary incontinence, gait disturbance and dementia~\cite{Eide2025,Malm2006}. The iNPH diagnosis is based on the American-European guidelines~\cite{Relkin2005-wf}, which not only consider the ventriculomegaly, however a hallmark of iNPH is enlarged cerebral ventricles associated with CSF flow disturbances~\cite{Ringstad2017,Eide2020}. This cohort is therefore expected to display the most consistent geometric variations compared to other patient groups. In this work, we only consider the effect of brain geometry on simulations of glymphatic function and accordingly consider this group as distinct. The remaining $70$ patients in the dataset consist of three main groups: 1) REF refers to patients who where not diagnosed with a CSF-related disorder following clinical work-up. These individuals may be considered as being close to healthy. 2) Cysts, refers to patients diagnosed with pineal or arachnoid cysts. 3) Other, refers to the remaining $26$ patients who are diagnosed with other CSF disorders such as Spontaneous Intracranial Hypotension (SIH) or Idiopathic Intracranial Hypertension (IIH). The non-iNPH cohort is therefore highly heterogeneous, however, from a purely geometrical perspective this group will, on the whole, be geometrically distinct from the iNPH group. 

\subsection{Implementation details}

For each MRI in the dataset, we build 3D digital meshes using the pipeline described in~\citet{mri2fem_1}. First, we perform a FreeSurfer segmentation~\cite{dale1999} of the brain, which allows us to extract the pial and ventricular surfaces. Volumetric meshes are created using SVMTK~\cite{svmtk}, generally consisting of on the order of $10^5$ nodes. Simulations are performed using finite element schemes in dolfinx (v. 0.9.0)~\cite{baratta_2023}.  For the two compartment model we use a second order Lagrange basis for the trial and test spaces, and for the MPET model we use first order Lagrange elements. In each case, this leads to a number of degrees of freedom on the order of $(1.5 \pm 0.5)\cdot 10^6$ for the high-fidelity model, with slight variations depending on the specific brain geometry. Image registration is performed using the framework introduced in~\citet{AVANTS2008}, implemented in the Advanced Normalization Tools (ANTS) toolbox~(v. 2.5.4)

For computational purposes, we consider only the cortical surface of the brain and the third and lateral ventricles. Segmentations of the fourth ventricle are generally unreliable when building finite element meshes because the channel from the fourth to the third ventricle will be only a few voxels wide, or disappear entirely. In the interest of time, we therefore choose to remove the cerebellum, as well as the fourth ventricle and the brain stem from the model to remove the need for manual correction of the segmentation.

%% file: 3_0_Results.tex
\section{Results} \label{sec:results}

In the following, we implement the approach outlined in Section~(\ref{sec:methods}), and apply it to the two models from Section~(\ref{sec:Models}). We choose two target geometries: One patient in the REF group, with no known CSF disorder or neurological disease, and another patient diagnosed with iNPH displaying ventriculomegaly. The REF patient meanwhile displays ventricular morphology within a normal range and we therefore consider the brain of this patient as a close approximation of a healthy brain. For each target subject, we measure the average approximation error incurred by mapping solutions from another geometry, as well as the singular values and reconstruction error due to the reduced approximation in terms of the relative $L_2$ error. In each case, we investigate the effect of considering the entire dataset, as well as when considering only the brains of iNPH or non-iNPH patients. In this regard, we consider the non-iNPH patient brains as representative of typical brain geometric variations, and the iNPH patient brains as representative of a class of brain geometries with enlarged ventricles.  

\input{3_1_Diffusion_example}
\input{3_2_MPET_example}

%% file: 3_1_Diffusion_example.tex
\subsection{Two-compartment model of tracer distribution} \label{sec:twocomp}

\begin{figure*}[t]
    \centering
    \begin{subfigure}{0.48\linewidth}
        \centering
        \includegraphics[width=\linewidth]{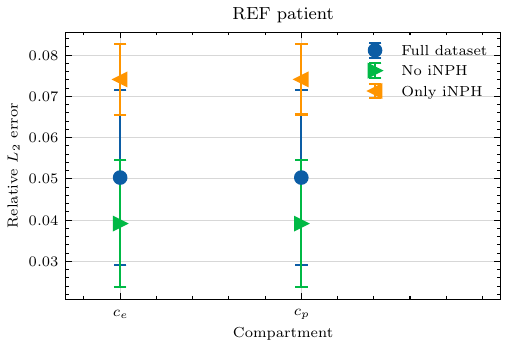}
        \caption{}
        \label{fig:twocomp_err_control}
    \end{subfigure}
    \begin{subfigure}{0.48\linewidth}
        \centering
        \includegraphics[width=\linewidth]{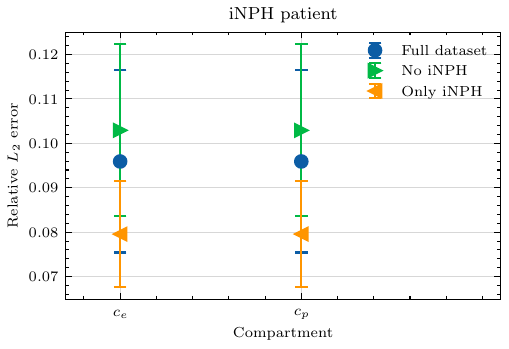}
        \caption{}
        \label{fig:twocomp_err_iNPH}
    \end{subfigure}
    \caption{Relative error in the $L_2$ norm on different subsets of the full dataset. We show results for (a) a REF patient and (b) a patient diagnosed with iNPH.}
    \label{fig:twocomp_err_L2}
\end{figure*}
 
\subsubsection{Approximation by any other geometry} \label{sec:twocomp_approx}

The relative error due to mapping solutions from one brain geometry to another appears to depend strongly on the target patient group when applied to Eq. (\ref{eq:twocomp}), as seen in Figure~(\ref{fig:twocomp_err_L2}). For the patient in the REF group, in Figure~(\ref{fig:twocomp_err_control}), the error is lower when considering only the brains of patients who are not diagnosed with iNPH. A similar, but opposite trend can be observed for the iNPH patient target brain, as seen in Figure~(\ref{fig:twocomp_err_iNPH}). On this target geometry, the relative error is also consistently larger across all dataset subdivisions. 

\subsubsection{Reduced order modeling}

\begin{figure*}[t]
    \centering
    \begin{subfigure}{0.48\linewidth}
        \centering
        \includegraphics[width=\linewidth]{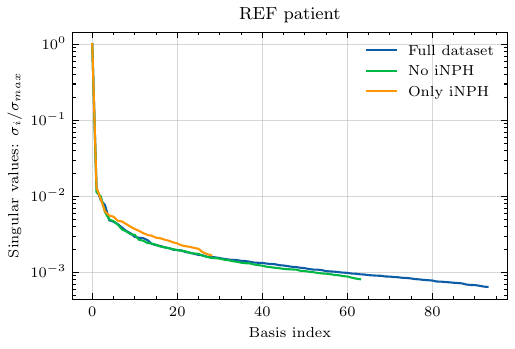}
        \caption{}
        \label{fig:twocomp_svd_control}
    \end{subfigure}
    \begin{subfigure}{0.48\linewidth}
        \centering
        \includegraphics[width=\linewidth]{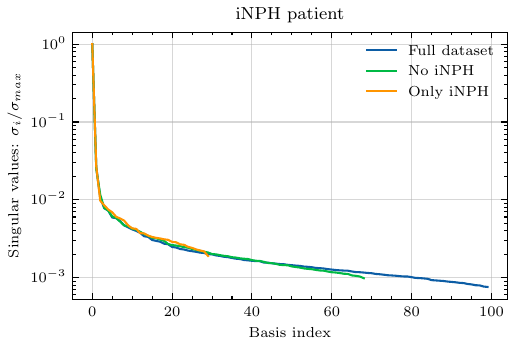}
        \caption{}
        \label{fig:twocomp_svd_iNPH}
    \end{subfigure}
    \caption{Normalized singular values as a function of the basis size. We show results for (a) a REF patient and (b) a patient diagnosed with iNPH.}
    \label{fig:twocomp_svd}
\end{figure*}

\begin{figure*}[t]
    \centering
    \begin{subfigure}{\linewidth}
        \centering
        \includegraphics[width=\linewidth]{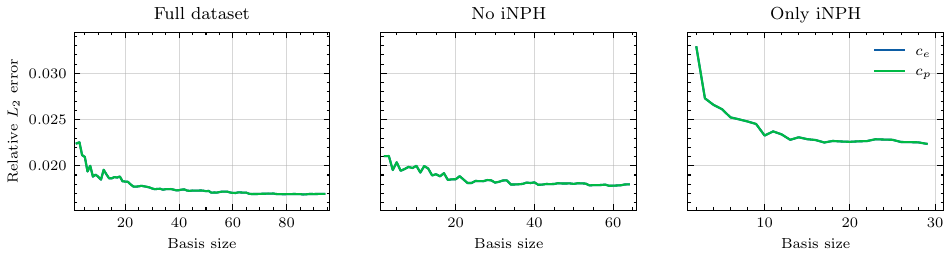}
        \caption{}
        \label{fig:twocomp_red_err_control}
    \end{subfigure}
    \begin{subfigure}{\linewidth}
        \centering
        \includegraphics[width=\linewidth]{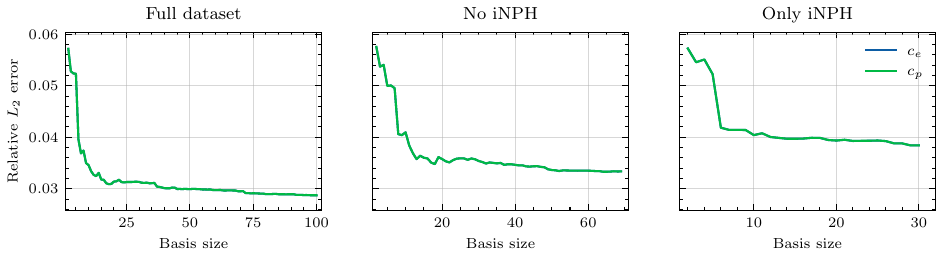}
        \caption{}
        \label{fig:twocomp_red_err_iNPH}
    \end{subfigure}
    \caption{Relative error in the $L_2$ norm of the reduced solution compared to the high-fidelity solution. We show results for (a) a REF patient and (b) a patient diagnosed with iNPH.}
    \label{fig:twocomp_red_err_L2}
\end{figure*}

On both target brain geometries, the singular values appear to decay by $3$ orders of magnitude from the first singular value, to the size of the dataset, as seen Figure (\ref{fig:twocomp_svd}). This decay is consistent across dataset subdivisions. Neither the REF, nor the iNPH patient display any clear advantage due to including only brains from the same patient group. This trend is further confirmed by Figure (\ref{fig:twocomp_red_err_L2}), where using the full dataset leads to lower error on both target brains. The minor exception to this rule is the middle plot of Figure (\ref{fig:twocomp_red_err_control}), where we notice that on the REF patient brain, excluding the patients diagnosed with iNPH appears to yield comparable results to using the entire dataset. 

\begin{figure*}[t]
    \centering
    \includegraphics[width=\linewidth]{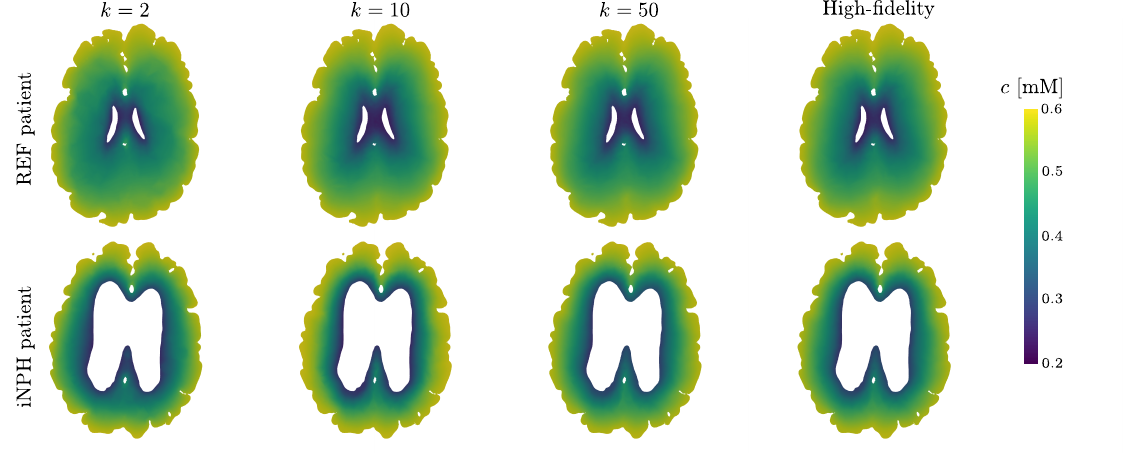}
    \caption{
    Comparison of concentration in the extracellular compartment $c_e$ using $k=2,10,50$ bases in the reduced solution for both the REF patient and the iNPH patient. We plot the high-fidelity solution for reference, and note the consistently strong similarity with the reduced solutions.
    }
    \label{fig:twocomp_ce_example}
\end{figure*}

In Figure~(\ref{fig:twocomp_ce_example}), we plot an example of an axial slice showing the ECS concentration computed from three reduced models with different numbers of bases $k$, compared to the high-fidelity model. We notice that even in the $k=2$ case, the reduced solution is generally very similar to the high-fidelity solution both for the REF patient brain and the iNPH patient brain. 

%% file: 3_2_MPET_example.tex
\subsection{Multi-compartment poro-elastic model (MPET)} \label{sec:MPET}

\begin{figure*}[t]
    \centering
    \begin{subfigure}{0.48\linewidth}
        \centering
        \includegraphics[width=\linewidth]{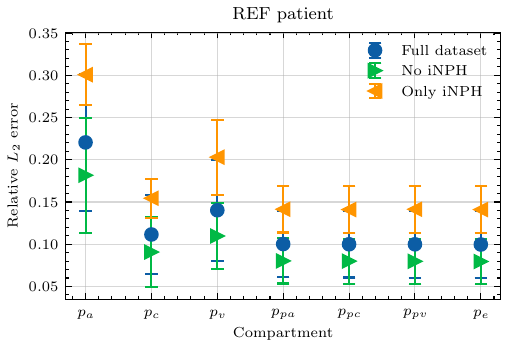}
        \caption{}
        \label{fig:MPET_err_control}
    \end{subfigure}
    \begin{subfigure}{0.48\linewidth}
        \centering
        \includegraphics[width=\linewidth]{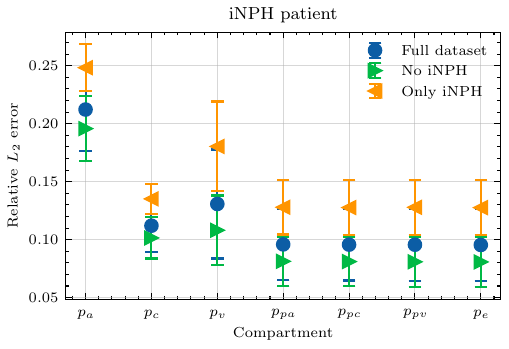}
        \caption{}
        \label{fig:MPET_err_iNPH}
    \end{subfigure}
    \caption{Relative error in the $L_2$ norm on different subset of the full dataset. We show results for (a) a REF patient and (b) a patient diagnosed with iNPH.}
    \label{fig:MPET_err_L2}
\end{figure*}

\subsubsection{Approximation by any other geometry}

When solving Eq.~(\ref{eq:MPET}), there does not appear to be any advantage gained from only mapping solutions from the same subject group, as shown in Figure~(\ref{fig:MPET_err_L2}). In particular, the relative error is larger when mapping solutions from brains of patients with iNPH in all compartments and on both target geometries. Indeed, in all cases, the relative error appears to be lower when considering only the brains of patients in the dataset who are not diagnosed with iNPH.

\subsubsection{Reduced order modeling}

\begin{figure*}[t]
    \centering
    \begin{subfigure}{0.48\linewidth}
        \centering
        \includegraphics[width=\linewidth]{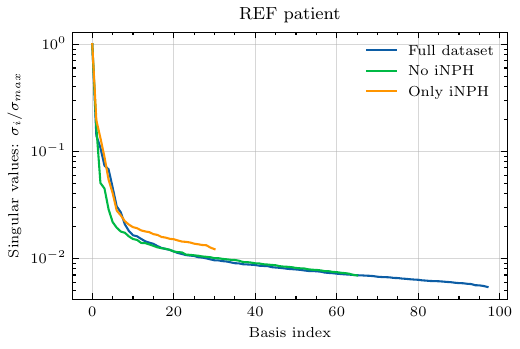}
        \caption{}
        \label{fig:MPET_svd_control}
    \end{subfigure}
    \begin{subfigure}{0.48\linewidth}
        \centering
        \includegraphics[width=\linewidth]{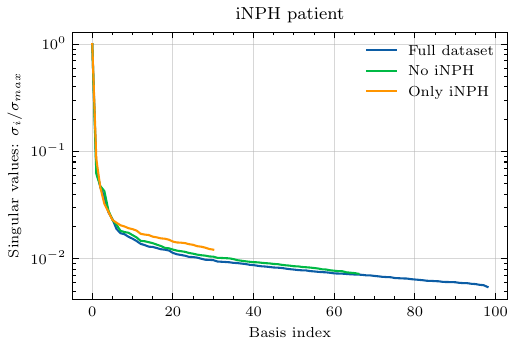}
        \caption{}
        \label{fig:MPET_svd_iNPH}
    \end{subfigure}
    \caption{Normalized singular values as a function of the basis size. We show results for (a) a REF patient and (b) a patient diagnosed with iNPH.}
    \label{fig:MPET_svd}
\end{figure*}

\begin{figure*}[t]
    \centering
    \begin{subfigure}{\linewidth}
        \centering
        \includegraphics[width=\linewidth]{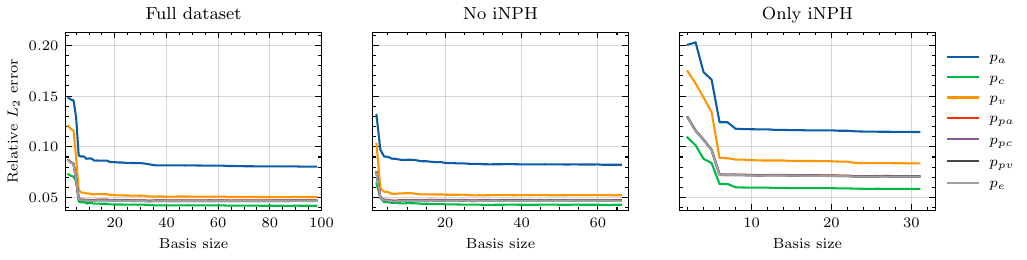}
        \caption{}
        \label{fig:MPET_red_err_control}
    \end{subfigure}
    \begin{subfigure}{\linewidth}
        \centering
        \includegraphics[width=\linewidth]{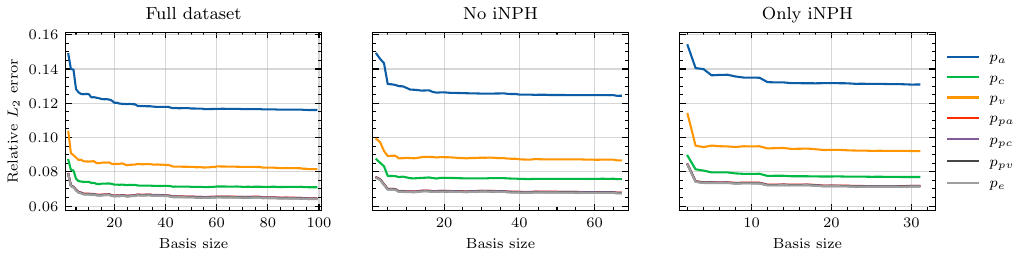}
        \caption{}
        \label{fig:MPET_red_err_iNPH}
    \end{subfigure}
    \caption{Relative error in the $L_2$ norm of the reduced solution compared to the high-fidelity solution. We show results for (a) a REF patient and (b) a patient diagnosed with iNPH.}
    \label{fig:MPET_red_err_L2}
\end{figure*}

On the MPET problem, the singular values appear to decay by $2$ orders of magnitude compared to the largest singular value when using the entire dataset of brain geometries, as seen in Figure~(\ref{fig:MPET_svd}). We also notice that the rate of decay on both target brains is slightly slower when using only iNPH brains, thus reflecting the results in Figure~(\ref{fig:MPET_err_L2}). Overall, there are no significant differences between the singular value spectra of the REF and iNPH target patients.   

The relative error due to the reduced basis approximation, shown in Figure~(\ref{fig:MPET_red_err_L2}), on the MPET problem follows a similar trend as the two-compartment problem in Figure~(\ref{fig:twocomp_red_err_L2}). In particular, we notice that on the REF target patient, seen in Figure~(\ref{fig:MPET_red_err_control}), there is effectively only a minor difference in performance between using the entire dataset (left) and when considering only the non-iNPH patients (middle). Only including brains from iNPH patients in the snapshot matrix leads to significantly larger errors on all compartments. We notice a similar trend on the iNPH target patient, in Figure~(\ref{fig:MPET_red_err_iNPH}), although the gaps between the dataset partitions appear less significant. In this case, it appears advantageous to use the entire dataset, and thus include the brains of patients with iNPH. The results in Figure~(\ref{fig:MPET_red_err_L2}) also display significant differences in the errors on the various compartments in the MPET model. In all cases, the arterial compartment is the clear outlier, leading to significantly larger errors than other fields. 

%% file: 4_Discussion.tex
\section{Discussion} \label{sec:Discussion}

\subsection{Approximation by any other geometry}

As a first step towards reduced order modeling, we evaluate the error of each solution, deformed from another geometry, in Figures~(\ref{fig:twocomp_err_L2}) and (\ref{fig:MPET_err_L2}). We notice that the errors are generally significantly higher in the $7$-compartment pressure model, with average errors ranging from $8-20$\% when excluding iNPH patients on both target subjects. Meanwhile in the two-compartment tracer model, the average error when considering only patients from the same cohort is $4$\% on the REF target patient and $8$\% on the iNPH target patient. The arterial compartment of the $7$-compartment model in Figure~(\ref{fig:MPET_err_L2}) is especially difficult to approximate using solutions from other geometries, largely as a result of the boundary conditions. At the cortical surface, the arterial compartment has a pure Neumann condition on the pial surface, with a substantial influx due to $Q_{in}$. As the pial surface is a folded sheet with relatively tight folds, this causes an artificial build-up of pressure in small areas. The solution, therefore, depends too strongly on the shape of the cortical folds, which image registration is not able to consistently capture, leading to an average $L_2$ error of more than $20$\% in most cases. We also note that the capillary compartment has a similar boundary condition to the arteries, but on the ventricular, rather than the pial, surface. As seen in Figure~(\ref{fig:MPET_err_L2}), the average error on this compartment is significantly lower, closer to $8$\%, as a result of the ventricular surface being more regular. Meanwhile, the boundary condition on the two-compartment problem leads to more smoothly varying solutions at the boundary, and as a result the errors in Figure~(\ref{fig:twocomp_err_L2}) are consistently lower than in Figure~(\ref{fig:MPET_err_L2}).

The other notable impact of the boundary conditions is the importance of considering the diagnosis of each patient when mapping solutions between brains. It is notable that in Figure~(\ref{fig:twocomp_err_L2}), the  relative error on the REF target patient is considerably lower when we only consider patients without iNPH, while on the iNPH patient the error is lower when we only consider patients with iNPH. In each case the average errors improve by close to $1$ percentage point compared to using the entire dataset. However, this trend is not at all present in Figure~(\ref{fig:MPET_err_L2}), where excluding the iNPH patients always leads to a lower relative error, generally lowering the error by more than $2$ percentage points compared to using the entire dataset. We notice that in Eq.~(\ref{eq:twocomp}) the Robin boundary condition is placed on the ventricular and pial surfaces, while in the $7$-compartment model all compartments, except the capillaries, have a homogeneous Neumann boundary condition on the ventricular surface. In the particular case of brains from patients with iNPH, the ventricular surface will inevitably be the most challenging aspect of image registration, as these anatomical features will vary considerably from the non-iNPH subjects. Consequently, applying the $c_{SAS}$ concentration on the ventricular surface in Eq.~(\ref{eq:twocomp}), means that matching this surface is critical to obtain a low error. In this case, it would therefore highly advantageous to use brains with the most similar morphology at the ventricles. Conversely, in the $7$-compartment model, the homogeneous Neumann condition at the ventricles leads to a lower dependence in the particular surface morphology. In this case, the increased challenge of registering MRI of patients with iNPH likely outweighs any advantage that is gained by the ventricles being more similar. Indeed, in Figure~(\ref{fig:MPET_err_L2}), we notice that including only the patients diagnosed with iNPH leads to a significantly larger increase in the relative error on the REF target patient than on the iNPH target patient.

\subsection{Reduced order modeling}

While Figures~(\ref{fig:twocomp_err_L2}) and (\ref{fig:MPET_err_L2}) show that there is some advantage to considering subsets of the full dataset when mapping solutions from one brain in the dataset to the target brain, this does not extend to reduced order modeling. For both problems, we build reduced bases based on the entire dataset, as well as when including only iNPH or non-iNPH patients. As evidenced by Figures~(\ref{fig:twocomp_svd}) and (\ref{fig:MPET_svd}) the singular value spectra of the three dataset variations are only minimally different. In some cases, it is advantageous to remove the iNPH patients from the dataset, but not importantly so. This is further verified by Figures~(\ref{fig:twocomp_red_err_L2}) and (\ref{fig:MPET_red_err_L2}). In all of these results, using the entire dataset yields superior or equivalent results to removing subjects diagnosed with iNPH, and in all cases considering only the iNPH individuals is worse, even when the target does itself have iNPH. The brain geometries which represent the most extreme morphological outliers will mostly lead to the optimization method not converging, and these cases are thus effectively removed from the dataset automatically. The cases where the optimization scheme succeeds are therefore more likely to be more morphologically representative and therefore pre-partitioning the dataset is counter-productive. Relying on the SVD to extract coherent basis vectors spanning the space of snapshot solutions is therefore a more consistent strategy than performing data triage based on a priori pathological information, under the assumption that the inter-brain mapping can be computed. 

When applied to the two-compartment tracer model, we obtain relative errors of less than $2\%$ on the REF target patient and close to $3\%$ on the iNPH case when using the largest reduced basis. Meanwhile, on the $7$-compartment model, errors are typically in the $9\%$ and $5\%$ range for the REF target patient and $12\%$ to $7\%$ for the iNPH target patient. In many biomedical applications, this error is smaller than the uncertainty due to measurement error. These results are also comparable to works on parameter estimation in the brain, such as~\citet{vinje2023human} or \citet{Valnes2020-jh}. We also note that the MRI data itself has a relatively low signal-to-noise ratio~\cite{Valnes2020-jh}, meaning that the error due to model order reduction would likely not be the largest factor when comparing with measurement data. For many applications involving parameter estimation from data, the results we obtain in this work will therefore likely be within a reasonable range.    

\subsection{Limitations}

As in any data-driven application analyzing medical data from humans, data availability is a fundamental limitation. While brain MRI from $101$ individuals is a relatively large number in a medical setting, typical problems involving computational design often leverage the capability of sampling several orders of magnitude more data points for the construction of a reduced representation.  In the context of geometry-based brain modeling we will never have access to this amount of data, and it is therefore critical that the brains that can be accessed are also sufficient to allow the construction of an efficient reduced basis. In an abstract sense, any brain geometry is inherently an extremely high-dimensional object with an almost infinite potential for individual variations. However, empirically, we know that brains display sufficiently predictable patterns to enable automated segmentation based purely on $T_1$-weighted MRI. In practice, the space of anatomically possible brains may therefore be significantly more compact than one may intuitively imagine.  We notice that in the case where the target brain is from REF group in the dataset, the error of the reduced order model, as seen in Figures~(\ref{fig:twocomp_red_err_control}) and (\ref{fig:MPET_red_err_control}) reaches a plateau after a very low number of bases. There is also only a negligible difference between the error obtained from the full dataset with $N \sim 100$ subjects and the non-iNPH subset of the dataset containing $\sim 70$ individuals. This indicates that adding MRI of more individuals to the dataset will likely have a minimal impact on the results, to the extent that we are not able to increase the number of available subjects by orders of magnitude. 

The other limitation of the present approach is that the brain-to-brain mappings are neither one-to-one, nor unique. We will inevitably not be able to map all points in a mesh built for one brain, to the exactly equivalent point on a mesh built for another brain. Solutions must therefore be extended from the deformed mesh, to the target mesh of interest. A key consequence of the mappings not being one-to-one is that solutions mapped to the target mesh will not necessarily span a conforming subspace of the relevant function space. In particular, we do not expect the boundary conditions to remain respected on the target geometry, as they were in the solution on the original geometry. As a consequence, the space spanned by the solutions mapped to the target geometry $\hat{V}_M = \text{span}(\hat{u}_1^h \dots \hat{u}_M^h)$ will not be a subspace of the appropriate Hilbert space $V_{target}$ for the variational problem on the target geometry. This is the main reason why the singular value decays in Figures~(\ref{fig:twocomp_svd}) and (\ref{fig:MPET_svd}) are relatively slow compared to what one might expect for the type of problems we solve in this work. A remedy to this issue may be to consider nonlinear dimensional reduction methods, such as a variational auto-encoder, which could potentially enforce the boundary condition more consistently than the SVD. However, this approach may prove challenging due to the relatively low amount of available geometries in the dataset. 

Finally, in this work we only consider PDEs which are independent of time for illustrative purposes. However, the present approach extends readily to time-dependent problems. The brain-to-brain mapping is a purely geometrical one, and remains valid as long as the brain geometries are independent of time.   

%% file: 5_Conclusion.tex
\section{Conclusion}

In this work, we propose a new method for mapping solutions on 3D models of brain geometries between meshes, using medical image registration. We implement an offline-online POD-based approach for building a reduced basis, and apply the method to a dataset of $101$ MRI of human individuals. The method is evaluated on two models of human glymphatic function: One problem modeling tracer transport in the brain~\cite{Riseth2024}, and a simplified $7$-compartment poro-elastic model~\cite{dreyer2024modeling} of the brain. For each of the two models, we consider one brain from a patient without any identified CSF disorder or neurological disease, considered close to healthy, and one brain from a patient diagnosed with iNPH. In each case, we perform high-fidelity simulations with degrees of freedom on the order of $10^6$, and evaluate the error reduced models of up to $100$ basis vectors. We are able to obtain reduced solutions with similar accuracy as state-of-the-art brain modeling methods. This work opens up a new avenue for leveraging datasets of brain geometries to build reduced models, an important first step towards patient-specific parameter estimation in glymphatic modeling.

%% file: 7_Supplementary_Declarations.tex
\section*{Declarations}

\subsection*{Funding}

K.A.M. and A.S. acknowledge funding by the European Research Council under grant 101141807 (aCleanBrain) and the national infrastructure for computational science in Norway, Sigma2, via grant NN9279K. This work was supported by the foundation Stiftelsen Kristian Gerhard Jebsen through its program for translational medical research.

\subsection*{Acknowledgements}
Computational experiments and data processing were performed on resources provided by Sigma2 — the National Infrastructure for High-Performance Computing and Data Storage in Norway, under grant grant NN9279K. A.S thanks Jørgen N. Riseth and Lars W. Dreyer for sharing code used for model implementation, and Lars M. Valnes for assistance in data preparation. 

\subsection*{Ethics and approvals}

Parts of the data presented in this has been presented in previous works on MRI-based assessment of human glymphatic function conducted at the University Hospital of Oslo~\citep{Ringstad2017,Ringstad2018-sv} in the years $2015-2016$. Collection of data analyzed for this study was approved by the Regional Committee for Medical and Health Research Ethics (REK) of Health Region South-East, Norway (2015/96), the Institutional Review Board of Oslo University Hospital (2015/1868) and the National Medicines Agency (15/04932-7), and conducted following the ethical standards of the Declaration of Helsinki of 1975 (revised in 1983). Study participants were included after written and oral informed consent. No new data was collected for the present work.

\subsection*{Competing interests}

The authors declare no competing interests.

\subsection*{Code and data availability}

Code for this paper, with scripts for registration and finite element simulations, is openly available from~\url{https://github.com/Erasdna/GROM}. The dataset used in this paper is not publicly available due to patient data privacy concerns. 

\subsection*{CRediT}

\textbf{Andreas Solheim}: Writing: Original draft, Review \& editing. Conceptualization, Methodology, Software, Investigation, Visualization. \textbf{Geir Ringstad}: Writing: Review \& editing, Data collection and curation. \textbf{Per Kristian Eide}: Writing: Review \& editing, Data collection and curation. \textbf{Kent-Andre Mardal}: Writing: Review \& editing, Conceptualization, Methodology, Supervision.

%% file: 0_Main.bbl
\begin{thebibliography}{47}
\expandafter\ifx\csname natexlab\endcsname\relax\def\natexlab#1{#1}\fi
\providecommand{\url}[1]{\texttt{#1}}
\providecommand{\href}[2]{#2}
\providecommand{\path}[1]{#1}
\providecommand{\DOIprefix}{doi:}
\providecommand{\ArXivprefix}{arXiv:}
\providecommand{\URLprefix}{URL: }
\providecommand{\Pubmedprefix}{pmid:}
\providecommand{\doi}[1]{\href{http://dx.doi.org/#1}{\path{#1}}}
\providecommand{\Pubmed}[1]{\href{pmid:#1}{\path{#1}}}
\providecommand{\bibinfo}[2]{#2}
\ifx\xfnm\relax \def\xfnm[#1]{\unskip,\space#1}\fi
\bibitem[{Iliff et~al.(2012)Iliff, Wang, Liao, Plog, Peng, Gundersen, Benveniste, Vates, Deane, Goldman, Nagelhus, and Nedergaard}]{Iliff2012}
\bibinfo{author}{J.~Iliff}, \bibinfo{author}{M.~Wang}, \bibinfo{author}{Y.~Liao}, \bibinfo{author}{B.~Plog}, \bibinfo{author}{W.~Peng}, \bibinfo{author}{G.~Gundersen}, \bibinfo{author}{H.~Benveniste}, \bibinfo{author}{G.~Vates}, \bibinfo{author}{R.~Deane}, \bibinfo{author}{S.~Goldman}, \bibinfo{author}{E.~Nagelhus}, \bibinfo{author}{M.~Nedergaard},
\newblock \bibinfo{title}{A paravascular pathway facilitates csf flow through the brain parenchyma and the clearance of interstitial solutes, including amyloid},
\newblock \bibinfo{journal}{Science translational medicine} \bibinfo{volume}{4} (\bibinfo{year}{2012}) \bibinfo{pages}{147ra111}. \DOIprefix\doi{10.1126/scitranslmed.3003748}.
\bibitem[{Louveau et~al.(2015)Louveau, Smirnov, Keyes, Eccles, Rouhani, Peske, Derecki, Castle, Mandell, Lee, Harris, and Kipnis}]{Louveau2015}
\bibinfo{author}{A.~Louveau}, \bibinfo{author}{I.~Smirnov}, \bibinfo{author}{T.~J. Keyes}, \bibinfo{author}{J.~D. Eccles}, \bibinfo{author}{S.~J. Rouhani}, \bibinfo{author}{J.~D. Peske}, \bibinfo{author}{N.~C. Derecki}, \bibinfo{author}{D.~Castle}, \bibinfo{author}{J.~W. Mandell}, \bibinfo{author}{K.~S. Lee}, \bibinfo{author}{T.~H. Harris}, \bibinfo{author}{J.~Kipnis},
\newblock \bibinfo{title}{Structural and functional features of central nervous system lymphatic vessels},
\newblock \bibinfo{journal}{Nature} \bibinfo{volume}{523} (\bibinfo{year}{2015}) \bibinfo{pages}{337--341}.
\bibitem[{Aspelund et~al.(2015)Aspelund, Antila, Proulx, Karlsen, Karaman, Detmar, Wiig, and Alitalo}]{Aspelund2015}
\bibinfo{author}{A.~Aspelund}, \bibinfo{author}{S.~Antila}, \bibinfo{author}{S.~T. Proulx}, \bibinfo{author}{T.~V. Karlsen}, \bibinfo{author}{S.~Karaman}, \bibinfo{author}{M.~Detmar}, \bibinfo{author}{H.~Wiig}, \bibinfo{author}{K.~Alitalo},
\newblock \bibinfo{title}{A dural lymphatic vascular system that drains brain interstitial fluid and macromolecules},
\newblock \bibinfo{journal}{Journal of Experimental Medicine} \bibinfo{volume}{212} (\bibinfo{year}{2015}) \bibinfo{pages}{991--999}. \URLprefix \url{https://doi.org/10.1084/jem.20142290}. \DOIprefix\doi{10.1084/jem.20142290}. \href{http://arxiv.org/abs/https://rupress.org/jem/article-pdf/212/7/991/1753793/jem\_20142290.pdf}{{\tt arXiv:https://rupress.org/jem/article-pdf/212/7/991/1753793/jem\_20142290.pdf}}.
\bibitem[{Xie et~al.(2013)Xie, Kang, Xu, Chen, Liao, Thiyagarajan, O’Donnell, Christensen, Nicholson, Iliff et~al.}]{xie2013sleep}
\bibinfo{author}{L.~Xie}, \bibinfo{author}{H.~Kang}, \bibinfo{author}{Q.~Xu}, \bibinfo{author}{M.~J. Chen}, \bibinfo{author}{Y.~Liao}, \bibinfo{author}{M.~Thiyagarajan}, \bibinfo{author}{J.~O’Donnell}, \bibinfo{author}{D.~J. Christensen}, \bibinfo{author}{C.~Nicholson}, \bibinfo{author}{J.~J. Iliff}, et~al.,
\newblock \bibinfo{title}{Sleep drives metabolite clearance from the adult brain},
\newblock \bibinfo{journal}{science} \bibinfo{volume}{342} (\bibinfo{year}{2013}) \bibinfo{pages}{373--377}.
\bibitem[{Bojarskaite et~al.(2023)Bojarskaite, Vallet, Bj{\o}rnstad, Gullestad~Binder, Cunen, Heuser, Kuchta, Mardal, and Enger}]{bojarskaite2023sleep}
\bibinfo{author}{L.~Bojarskaite}, \bibinfo{author}{A.~Vallet}, \bibinfo{author}{D.~M. Bj{\o}rnstad}, \bibinfo{author}{K.~M. Gullestad~Binder}, \bibinfo{author}{C.~Cunen}, \bibinfo{author}{K.~Heuser}, \bibinfo{author}{M.~Kuchta}, \bibinfo{author}{K.-A. Mardal}, \bibinfo{author}{R.~Enger},
\newblock \bibinfo{title}{Sleep cycle-dependent vascular dynamics in male mice and the predicted effects on perivascular cerebrospinal fluid flow and solute transport},
\newblock \bibinfo{journal}{Nature communications} \bibinfo{volume}{14} (\bibinfo{year}{2023}) \bibinfo{pages}{953}.
\bibitem[{Eide et~al.(2021)Eide, Vinje, Pripp, Mardal, and Ringstad}]{eide2021sleep}
\bibinfo{author}{P.~K. Eide}, \bibinfo{author}{V.~Vinje}, \bibinfo{author}{A.~H. Pripp}, \bibinfo{author}{K.-A. Mardal}, \bibinfo{author}{G.~Ringstad},
\newblock \bibinfo{title}{Sleep deprivation impairs molecular clearance from the human brain},
\newblock \bibinfo{journal}{Brain} \bibinfo{volume}{144} (\bibinfo{year}{2021}) \bibinfo{pages}{863--874}.
\bibitem[{Hardy and Selkoe(2002)}]{Hardy2002}
\bibinfo{author}{J.~Hardy}, \bibinfo{author}{D.~J. Selkoe},
\newblock \bibinfo{title}{The amyloid hypothesis of alzheimer's disease: Progress and problems on the road to therapeutics},
\newblock \bibinfo{journal}{Science} \bibinfo{volume}{297} (\bibinfo{year}{2002}) \bibinfo{pages}{353--356}. \URLprefix \url{https://www.science.org/doi/abs/10.1126/science.1072994}. \DOIprefix\doi{10.1126/science.1072994}. \href{http://arxiv.org/abs/https://www.science.org/doi/pdf/10.1126/science.1072994}{{\tt arXiv:https://www.science.org/doi/pdf/10.1126/science.1072994}}.
\bibitem[{Spillantini and Goedert(2013)}]{Spillantini2013}
\bibinfo{author}{M.~G. Spillantini}, \bibinfo{author}{M.~Goedert},
\newblock \bibinfo{title}{Tau pathology and neurodegeneration},
\newblock \bibinfo{journal}{Lancet Neurol} \bibinfo{volume}{12} (\bibinfo{year}{2013}) \bibinfo{pages}{609--622}.
\bibitem[{Goedert(2001)}]{Goedert2001}
\bibinfo{author}{M.~Goedert},
\newblock \bibinfo{title}{Alpha-synuclein and neurodegenerative diseases},
\newblock \bibinfo{journal}{Nature Reviews Neuroscience} \bibinfo{volume}{2} (\bibinfo{year}{2001}) \bibinfo{pages}{492--501}.
\bibitem[{Corti et~al.(2024)Corti, Bonizzoni, and Antonietti}]{corti2024structure}
\bibinfo{author}{M.~Corti}, \bibinfo{author}{F.~Bonizzoni}, \bibinfo{author}{P.~F. Antonietti},
\newblock \bibinfo{title}{Structure preserving polytopal discontinuous galerkin methods for the numerical modeling of neurodegenerative diseases},
\newblock \bibinfo{journal}{Journal of Scientific Computing} \bibinfo{volume}{100} (\bibinfo{year}{2024}) \bibinfo{pages}{39}.
\bibitem[{Dreyer et~al.(2024)Dreyer, Eklund, Rognes, Malm, Qvarlander, St{\o}verud, Mardal, and Vinje}]{dreyer2024modeling}
\bibinfo{author}{L.~W. Dreyer}, \bibinfo{author}{A.~Eklund}, \bibinfo{author}{M.~E. Rognes}, \bibinfo{author}{J.~Malm}, \bibinfo{author}{S.~Qvarlander}, \bibinfo{author}{K.-H. St{\o}verud}, \bibinfo{author}{K.-A. Mardal}, \bibinfo{author}{V.~Vinje},
\newblock \bibinfo{title}{Modeling csf circulation and the glymphatic system during infusion using subject specific intracranial pressures and brain geometries},
\newblock \bibinfo{journal}{Fluids and Barriers of the CNS} \bibinfo{volume}{21} (\bibinfo{year}{2024}) \bibinfo{pages}{82}.
\bibitem[{Fumagalli et~al.(2024)Fumagalli, Corti, Parolini, and Antonietti}]{fumagalli2024polytopal}
\bibinfo{author}{I.~Fumagalli}, \bibinfo{author}{M.~Corti}, \bibinfo{author}{N.~Parolini}, \bibinfo{author}{P.~F. Antonietti},
\newblock \bibinfo{title}{Polytopal discontinuous galerkin discretization of brain multiphysics flow dynamics},
\newblock \bibinfo{journal}{Journal of Computational Physics} \bibinfo{volume}{513} (\bibinfo{year}{2024}) \bibinfo{pages}{113115}.
\bibitem[{Hornkjøl et~al.(2022)Hornkjøl, Valnes, Ringstad, Rognes, Eide, Mardal, and Vinje}]{Hornkjol2022}
\bibinfo{author}{M.~Hornkjøl}, \bibinfo{author}{L.~M. Valnes}, \bibinfo{author}{G.~Ringstad}, \bibinfo{author}{M.~E. Rognes}, \bibinfo{author}{P.-K. Eide}, \bibinfo{author}{K.-A. Mardal}, \bibinfo{author}{V.~Vinje},
\newblock \bibinfo{title}{Csf circulation and dispersion yield rapid clearance from intracranial compartments},
\newblock \bibinfo{journal}{Frontiers in Bioengineering and Biotechnology} \bibinfo{volume}{10} (\bibinfo{year}{2022}). \URLprefix \url{https://www.frontiersin.org/journals/bioengineering-and-biotechnology/articles/10.3389/fbioe.2022.932469}. \DOIprefix\doi{10.3389/fbioe.2022.932469}.
\bibitem[{Johnson et~al.(2023)Johnson, Abdelmalik, Baidoo, Badachhape, Hughes, and Hossain}]{johnson2023image}
\bibinfo{author}{M.~J. Johnson}, \bibinfo{author}{M.~R. Abdelmalik}, \bibinfo{author}{F.~A. Baidoo}, \bibinfo{author}{A.~Badachhape}, \bibinfo{author}{T.~J. Hughes}, \bibinfo{author}{S.~S. Hossain},
\newblock \bibinfo{title}{Image-guided subject-specific modeling of glymphatic transport and amyloid deposition},
\newblock \bibinfo{journal}{Computer methods in applied mechanics and engineering} \bibinfo{volume}{417} (\bibinfo{year}{2023}) \bibinfo{pages}{116449}.
\bibitem[{Valnes et~al.(2020)Valnes, Mitusch, Ringstad, Eide, Funke, and Mardal}]{Valnes2020-jh}
\bibinfo{author}{L.~M. Valnes}, \bibinfo{author}{S.~K. Mitusch}, \bibinfo{author}{G.~Ringstad}, \bibinfo{author}{P.~K. Eide}, \bibinfo{author}{S.~W. Funke}, \bibinfo{author}{K.-A. Mardal},
\newblock \bibinfo{title}{Apparent diffusion coefficient estimates based on 24 hours tracer movement support glymphatic transport in human cerebral cortex},
\newblock \bibinfo{journal}{Scientific Reports} \bibinfo{volume}{10} (\bibinfo{year}{2020}) \bibinfo{pages}{9176}.
\bibitem[{Vardakis et~al.(2020)Vardakis, Chou, Guo, and Ventikos}]{vardakis2020exploring}
\bibinfo{author}{J.~C. Vardakis}, \bibinfo{author}{D.~Chou}, \bibinfo{author}{L.~Guo}, \bibinfo{author}{Y.~Ventikos},
\newblock \bibinfo{title}{Exploring neurodegenerative disorders using a novel integrated model of cerebral transport: Initial results},
\newblock \bibinfo{journal}{Proceedings of the Institution of Mechanical Engineers, Part H: Journal of Engineering in Medicine} \bibinfo{volume}{234} (\bibinfo{year}{2020}) \bibinfo{pages}{1223--1234}.
\bibitem[{Vinje et~al.(2023)Vinje, Zapf, Ringstad, Eide, Rognes, and Mardal}]{vinje2023human}
\bibinfo{author}{V.~Vinje}, \bibinfo{author}{B.~Zapf}, \bibinfo{author}{G.~Ringstad}, \bibinfo{author}{P.~K. Eide}, \bibinfo{author}{M.~E. Rognes}, \bibinfo{author}{K.-A. Mardal},
\newblock \bibinfo{title}{Human brain solute transport quantified by glymphatic mri-informed biophysics during sleep and sleep deprivation},
\newblock \bibinfo{journal}{Fluids and Barriers of the CNS} \bibinfo{volume}{20} (\bibinfo{year}{2023}) \bibinfo{pages}{62}.
\bibitem[{Chinesta et~al.(2016)Chinesta, Huerta, Rozza, and Willcox}]{chinesta2016}
\bibinfo{author}{F.~Chinesta}, \bibinfo{author}{A.~Huerta}, \bibinfo{author}{G.~Rozza}, \bibinfo{author}{K.~Willcox},
\newblock \bibinfo{title}{Model order reduction},
\newblock \bibinfo{journal}{Encyclopedia of computational mechanics}  (\bibinfo{year}{2016}) \bibinfo{pages}{1--36}.
\bibitem[{Hesthaven et~al.(2016)Hesthaven, Rozza, Stamm et~al.}]{hesthaven2016certified}
\bibinfo{author}{J.~S. Hesthaven}, \bibinfo{author}{G.~Rozza}, \bibinfo{author}{B.~Stamm}, et~al., \bibinfo{title}{Certified reduced basis methods for parametrized partial differential equations}, volume \bibinfo{volume}{590}, \bibinfo{publisher}{Springer}, \bibinfo{year}{2016}.
\bibitem[{Mardal et~al.(2022)Mardal, Rognes, Thompson, and Valnes}]{mri2fem_1}
\bibinfo{author}{K.-A. Mardal}, \bibinfo{author}{M.~Rognes}, \bibinfo{author}{T.~Thompson}, \bibinfo{author}{L.~Valnes}, \bibinfo{title}{Mathematical Modeling of the Human Brain: From Magnetic Resonance Images to Finite Element Simulation}, \bibinfo{publisher}{Springer}, \bibinfo{year}{2022}. \DOIprefix\doi{10.1007/978-3-030-95136-8}.
\bibitem[{Dale et~al.(1999)Dale, Fischl, and Sereno}]{dale1999}
\bibinfo{author}{A.~M. Dale}, \bibinfo{author}{B.~Fischl}, \bibinfo{author}{M.~I. Sereno},
\newblock \bibinfo{title}{Cortical surface-based analysis: I. segmentation and surface reconstruction},
\newblock \bibinfo{journal}{NeuroImage} \bibinfo{volume}{9} (\bibinfo{year}{1999}) \bibinfo{pages}{179--194}. \URLprefix \url{https://www.sciencedirect.com/science/article/pii/S1053811998903950}. \DOIprefix\doi{https://doi.org/10.1006/nimg.1998.0395}.
\bibitem[{Logg et~al.(2012)Logg, Mardal, and Wells}]{logg2012automated}
\bibinfo{author}{A.~Logg}, \bibinfo{author}{K.-A. Mardal}, \bibinfo{author}{G.~Wells}, \bibinfo{title}{Automated solution of differential equations by the finite element method: The FEniCS book}, volume~\bibinfo{volume}{84}, \bibinfo{publisher}{Springer Science \& Business Media}, \bibinfo{year}{2012}.
\bibitem[{Fischer and Modersitzki(2003)}]{fischer2003flirt}
\bibinfo{author}{B.~Fischer}, \bibinfo{author}{J.~Modersitzki},
\newblock \bibinfo{title}{Flirt: A flexible image registration toolbox},
\newblock in: \bibinfo{booktitle}{Biomedical Image Registration: Second InternationalWorkshop, WBIR 2003, Philadelphia, PA, USA, June 23-24, 2003. Revised Papers 2}, \bibinfo{organization}{Springer}, \bibinfo{year}{2003}, pp. \bibinfo{pages}{261--270}.
\bibitem[{Andersson et~al.(2008)Andersson, Smith, and Jenkinson}]{andersson2008fnirt}
\bibinfo{author}{J.~Andersson}, \bibinfo{author}{S.~Smith}, \bibinfo{author}{M.~Jenkinson},
\newblock \bibinfo{title}{Fnirt-fmrib’s non-linear image registration tool},
\newblock \bibinfo{journal}{Human Brain Mapping} \bibinfo{volume}{2008} (\bibinfo{year}{2008}).
\bibitem[{Ashburner(2007)}]{ASHBURNER2007}
\bibinfo{author}{J.~Ashburner},
\newblock \bibinfo{title}{A fast diffeomorphic image registration algorithm},
\newblock \bibinfo{journal}{NeuroImage} \bibinfo{volume}{38} (\bibinfo{year}{2007}) \bibinfo{pages}{95--113}. \URLprefix \url{https://www.sciencedirect.com/science/article/pii/S1053811907005848}. \DOIprefix\doi{https://doi.org/10.1016/j.neuroimage.2007.07.007}.
\bibitem[{Ashburner and Friston(2011)}]{ASHBURNER2011}
\bibinfo{author}{J.~Ashburner}, \bibinfo{author}{K.~J. Friston},
\newblock \bibinfo{title}{Diffeomorphic registration using geodesic shooting and gauss–newton optimisation},
\newblock \bibinfo{journal}{NeuroImage} \bibinfo{volume}{55} (\bibinfo{year}{2011}) \bibinfo{pages}{954--967}. \URLprefix \url{https://www.sciencedirect.com/science/article/pii/S1053811910016496}. \DOIprefix\doi{https://doi.org/10.1016/j.neuroimage.2010.12.049}.
\bibitem[{Hoffmann et~al.(2022)Hoffmann, Billot, Greve, Iglesias, Fischl, and Dalca}]{Hoffmann2022}
\bibinfo{author}{M.~Hoffmann}, \bibinfo{author}{B.~Billot}, \bibinfo{author}{D.~N. Greve}, \bibinfo{author}{J.~E. Iglesias}, \bibinfo{author}{B.~Fischl}, \bibinfo{author}{A.~V. Dalca},
\newblock \bibinfo{title}{Synthmorph: Learning contrast-invariant registration without acquired images},
\newblock \bibinfo{journal}{IEEE Transactions on Medical Imaging} \bibinfo{volume}{41} (\bibinfo{year}{2022}) \bibinfo{pages}{543--558}. \DOIprefix\doi{10.1109/TMI.2021.3116879}.
\bibitem[{Postelnicu et~al.(2009)Postelnicu, Zollei, and Fischl}]{Postelnicu2009}
\bibinfo{author}{G.~Postelnicu}, \bibinfo{author}{L.~Zollei}, \bibinfo{author}{B.~Fischl},
\newblock \bibinfo{title}{Combined volumetric and surface registration},
\newblock \bibinfo{journal}{IEEE Transactions on Medical Imaging} \bibinfo{volume}{28} (\bibinfo{year}{2009}) \bibinfo{pages}{508--522}. \DOIprefix\doi{10.1109/TMI.2008.2004426}.
\bibitem[{Avants et~al.(2008)Avants, Epstein, Grossman, and Gee}]{AVANTS2008}
\bibinfo{author}{B.~Avants}, \bibinfo{author}{C.~Epstein}, \bibinfo{author}{M.~Grossman}, \bibinfo{author}{J.~Gee},
\newblock \bibinfo{title}{Symmetric diffeomorphic image registration with cross-correlation: Evaluating automated labeling of elderly and neurodegenerative brain},
\newblock \bibinfo{journal}{Medical Image Analysis} \bibinfo{volume}{12} (\bibinfo{year}{2008}) \bibinfo{pages}{26--41}. \URLprefix \url{https://www.sciencedirect.com/science/article/pii/S1361841507000606}. \DOIprefix\doi{https://doi.org/10.1016/j.media.2007.06.004}, \bibinfo{note}{special Issue on The Third International Workshop on Biomedical Image Registration – WBIR 2006}.
\bibitem[{Modat et~al.(2014)Modat, Cash, Daga, Winston, Duncan, and Ourselin}]{Modat2014-er}
\bibinfo{author}{M.~Modat}, \bibinfo{author}{D.~M. Cash}, \bibinfo{author}{P.~Daga}, \bibinfo{author}{G.~P. Winston}, \bibinfo{author}{J.~S. Duncan}, \bibinfo{author}{S.~Ourselin},
\newblock \bibinfo{title}{Global image registration using a symmetric block-matching approach},
\newblock \bibinfo{journal}{J Med Imaging (Bellingham)} \bibinfo{volume}{1} (\bibinfo{year}{2014}) \bibinfo{pages}{024003}.
\bibitem[{Klein et~al.(2009)Klein, Andersson, Ardekani, Ashburner, Avants, Chiang, Christensen, Collins, Gee, Hellier, Song, Jenkinson, Lepage, Rueckert, Thompson, Vercauteren, Woods, Mann, and Parsey}]{Klein2009}
\bibinfo{author}{A.~Klein}, \bibinfo{author}{J.~Andersson}, \bibinfo{author}{B.~A. Ardekani}, \bibinfo{author}{J.~Ashburner}, \bibinfo{author}{B.~Avants}, \bibinfo{author}{M.-C. Chiang}, \bibinfo{author}{G.~E. Christensen}, \bibinfo{author}{D.~L. Collins}, \bibinfo{author}{J.~Gee}, \bibinfo{author}{P.~Hellier}, \bibinfo{author}{J.~H. Song}, \bibinfo{author}{M.~Jenkinson}, \bibinfo{author}{C.~Lepage}, \bibinfo{author}{D.~Rueckert}, \bibinfo{author}{P.~Thompson}, \bibinfo{author}{T.~Vercauteren}, \bibinfo{author}{R.~P. Woods}, \bibinfo{author}{J.~J. Mann}, \bibinfo{author}{R.~V. Parsey},
\newblock \bibinfo{title}{Evaluation of 14 nonlinear deformation algorithms applied to human brain {MRI} registration},
\newblock \bibinfo{journal}{Neuroimage} \bibinfo{volume}{46} (\bibinfo{year}{2009}) \bibinfo{pages}{786--802}.
\bibitem[{Klein et~al.(2010)Klein, Ghosh, Avants, Yeo, Fischl, Ardekani, Gee, Mann, and Parsey}]{Klein2010}
\bibinfo{author}{A.~Klein}, \bibinfo{author}{S.~S. Ghosh}, \bibinfo{author}{B.~Avants}, \bibinfo{author}{B.~T.~T. Yeo}, \bibinfo{author}{B.~Fischl}, \bibinfo{author}{B.~Ardekani}, \bibinfo{author}{J.~C. Gee}, \bibinfo{author}{J.~J. Mann}, \bibinfo{author}{R.~V. Parsey},
\newblock \bibinfo{title}{Evaluation of volume-based and surface-based brain image registration methods},
\newblock \bibinfo{journal}{Neuroimage} \bibinfo{volume}{51} (\bibinfo{year}{2010}) \bibinfo{pages}{214--220}.
\bibitem[{Eide(2025)}]{Eide2025}
\bibinfo{author}{P.~K. Eide},
\newblock \bibinfo{title}{Adult hydrocephalus and the glymphatic system},
\newblock \bibinfo{journal}{Neurosurgery Clinics of North America} \bibinfo{volume}{36} (\bibinfo{year}{2025}) \bibinfo{pages}{127--140}. \URLprefix \url{https://www.sciencedirect.com/science/article/pii/S1042368024000895}. \DOIprefix\doi{https://doi.org/10.1016/j.nec.2024.11.002}, \bibinfo{note}{adult Hydrocephalus and Intracranial Pressure Disorders}.
\bibitem[{Malm and Eklund(2006)}]{Malm2006}
\bibinfo{author}{J.~Malm}, \bibinfo{author}{A.~Eklund},
\newblock \bibinfo{title}{Idiopathic normal pressure hydrocephalus},
\newblock \bibinfo{journal}{Practical Neurology} \bibinfo{volume}{6} (\bibinfo{year}{2006}) \bibinfo{pages}{14--27}. \URLprefix \url{https://pn.bmj.com/content/6/1/14}. \DOIprefix\doi{10.1136/jnnp.2006.088351}. \href{http://arxiv.org/abs/https://pn.bmj.com/content/6/1/14.full.pdf}{{\tt arXiv:https://pn.bmj.com/content/6/1/14.full.pdf}}.
\bibitem[{Sirovich(1991)}]{Sirovich_1991}
\bibinfo{author}{L.~Sirovich},
\newblock \bibinfo{title}{Analysis of turbulent flows by means of the empirical eigenfunctions},
\newblock \bibinfo{journal}{Fluid Dynamics Research} \bibinfo{volume}{8} (\bibinfo{year}{1991}) \bibinfo{pages}{85}. \URLprefix \url{https://dx.doi.org/10.1016/0169-5983(91)90033-F}. \DOIprefix\doi{10.1016/0169-5983(91)90033-F}.
\bibitem[{Volkwein(2011)}]{volkwein2011model}
\bibinfo{author}{S.~Volkwein},
\newblock \bibinfo{title}{Model reduction using proper orthogonal decomposition},
\newblock \bibinfo{journal}{Lecture Notes, Institute of Mathematics and Scientific Computing, University of Graz} \bibinfo{volume}{\href{http://www. uni-graz. at/imawww/volkwein/POD. pdf}{1025}} (\bibinfo{year}{2011}).
\bibitem[{Eckart and Young(1936)}]{Eckart1936}
\bibinfo{author}{C.~Eckart}, \bibinfo{author}{G.~M. Young},
\newblock \bibinfo{title}{The approximation of one matrix by another of lower rank},
\newblock \bibinfo{journal}{Psychometrika} \bibinfo{volume}{1} (\bibinfo{year}{1936}) \bibinfo{pages}{211--218}. \URLprefix \url{https://api.semanticscholar.org/CorpusID:10163399}.
\bibitem[{Riseth et~al.(2024)Riseth, Koch, and Mardal}]{Riseth2024}
\bibinfo{author}{J.~N. Riseth}, \bibinfo{author}{T.~Koch}, \bibinfo{author}{K.-A. Mardal}, \bibinfo{title}{Two-compartment modeling of tracer transport in the brain}, \bibinfo{publisher}{Springer}, \bibinfo{address}{New York, USA}, \bibinfo{year}{2024}.
\bibitem[{Tully and Ventikos(2011)}]{TULLY_VENTIKOS_2011}
\bibinfo{author}{B.~Tully}, \bibinfo{author}{Y.~Ventikos},
\newblock \bibinfo{title}{Cerebral water transport using multiple-network poroelastic theory: application to normal pressure hydrocephalus},
\newblock \bibinfo{journal}{Journal of Fluid Mechanics} \bibinfo{volume}{667} (\bibinfo{year}{2011}) \bibinfo{pages}{188–215}. \DOIprefix\doi{10.1017/S0022112010004428}.
\bibitem[{Lee et~al.(2019)Lee, Piersanti, Mardal, and Rognes}]{lee2019mixed}
\bibinfo{author}{J.~J. Lee}, \bibinfo{author}{E.~Piersanti}, \bibinfo{author}{K.-A. Mardal}, \bibinfo{author}{M.~E. Rognes},
\newblock \bibinfo{title}{A mixed finite element method for nearly incompressible multiple-network poroelasticity},
\newblock \bibinfo{journal}{SIAM journal on scientific computing} \bibinfo{volume}{41} (\bibinfo{year}{2019}) \bibinfo{pages}{A722--A747}.
\bibitem[{Guo et~al.(2019)Guo, Li, Lyu, Mei, Vardakis, Chen, Han, Lou, and Ventikos}]{Guo2019-ss}
\bibinfo{author}{L.~Guo}, \bibinfo{author}{Z.~Li}, \bibinfo{author}{J.~Lyu}, \bibinfo{author}{Y.~Mei}, \bibinfo{author}{J.~C. Vardakis}, \bibinfo{author}{D.~Chen}, \bibinfo{author}{C.~Han}, \bibinfo{author}{X.~Lou}, \bibinfo{author}{Y.~Ventikos},
\newblock \bibinfo{title}{On the validation of a {Multiple-Network} poroelastic model using arterial spin labeling {MRI} data},
\newblock \bibinfo{journal}{Front Comput Neurosci} \bibinfo{volume}{13} (\bibinfo{year}{2019}) \bibinfo{pages}{60}.
\bibitem[{Ringstad et~al.(2017)Ringstad, Vatnehol, and Eide}]{Ringstad2017}
\bibinfo{author}{G.~Ringstad}, \bibinfo{author}{S.~A.~S. Vatnehol}, \bibinfo{author}{P.~K. Eide},
\newblock \bibinfo{title}{Glymphatic mri in idiopathic normal pressure hydrocephalus},
\newblock \bibinfo{journal}{Brain} \bibinfo{volume}{140} (\bibinfo{year}{2017}) \bibinfo{pages}{2691--2705}. \URLprefix \url{https://doi.org/10.1093/brain/awx191}. \DOIprefix\doi{10.1093/brain/awx191}. \href{http://arxiv.org/abs/https://academic.oup.com/brain/article-pdf/140/10/2691/24174132/awx191.pdf}{{\tt arXiv:https://academic.oup.com/brain/article-pdf/140/10/2691/24174132/awx191.pdf}}.
\bibitem[{Ringstad et~al.(2018)Ringstad, Valnes, Dale, Pripp, Vatnehol, Emblem, Mardal, and Eide}]{Ringstad2018-sv}
\bibinfo{author}{G.~Ringstad}, \bibinfo{author}{L.~M. Valnes}, \bibinfo{author}{A.~M. Dale}, \bibinfo{author}{A.~H. Pripp}, \bibinfo{author}{S.-A.~S. Vatnehol}, \bibinfo{author}{K.~E. Emblem}, \bibinfo{author}{K.-A. Mardal}, \bibinfo{author}{P.~K. Eide},
\newblock \bibinfo{title}{Brain-wide glymphatic enhancement and clearance in humans assessed with {MRI}},
\newblock \bibinfo{journal}{JCI Insight} \bibinfo{volume}{3} (\bibinfo{year}{2018}).
\bibitem[{Relkin et~al.(2005)Relkin, Marmarou, Klinge, Bergsneider, and Black}]{Relkin2005-wf}
\bibinfo{author}{N.~Relkin}, \bibinfo{author}{A.~Marmarou}, \bibinfo{author}{P.~Klinge}, \bibinfo{author}{M.~Bergsneider}, \bibinfo{author}{P.~M. Black},
\newblock \bibinfo{title}{Diagnosing idiopathic normal-pressure hydrocephalus},
\newblock \bibinfo{journal}{Neurosurgery} \bibinfo{volume}{57} (\bibinfo{year}{2005}).
\bibitem[{Eide et~al.(2020)Eide, Pripp, and Ringstad}]{Eide2020}
\bibinfo{author}{P.~K. Eide}, \bibinfo{author}{A.~H. Pripp}, \bibinfo{author}{G.~Ringstad},
\newblock \bibinfo{title}{Magnetic resonance imaging biomarkers of cerebrospinal fluid tracer dynamics in idiopathic normal pressure hydrocephalus},
\newblock \bibinfo{journal}{Brain Communications} \bibinfo{volume}{2} (\bibinfo{year}{2020}) \bibinfo{pages}{fcaa187}. \URLprefix \url{https://doi.org/10.1093/braincomms/fcaa187}. \DOIprefix\doi{10.1093/braincomms/fcaa187}. \href{http://arxiv.org/abs/https://academic.oup.com/braincomms/article-pdf/2/2/fcaa187/35055454/fcaa187.pdf}{{\tt arXiv:https://academic.oup.com/braincomms/article-pdf/2/2/fcaa187/35055454/fcaa187.pdf}}.
\bibitem[{Valnes and Schreiner(2021)}]{svmtk}
\bibinfo{author}{L.~Valnes}, \bibinfo{author}{J.~Schreiner}, \bibinfo{title}{Surface volume meshing toolkit (svmtk)}, \bibinfo{howpublished}{\url{https://github.com/SVMTK/SVMTK}}, \bibinfo{year}{2021}.
\bibitem[{Baratta et~al.(2023)Baratta, Dean, Dokken, Habera, Hale, Richardson, Rognes, Scroggs, Sime, and Wells}]{baratta_2023}
\bibinfo{author}{I.~A. Baratta}, \bibinfo{author}{J.~P. Dean}, \bibinfo{author}{J.~S. Dokken}, \bibinfo{author}{M.~Habera}, \bibinfo{author}{J.~S. Hale}, \bibinfo{author}{C.~N. Richardson}, \bibinfo{author}{M.~E. Rognes}, \bibinfo{author}{M.~W. Scroggs}, \bibinfo{author}{N.~Sime}, \bibinfo{author}{G.~N. Wells}, \bibinfo{title}{Dolfinx: The next generation fenics problem solving environment}, \bibinfo{year}{2023}. \URLprefix \url{https://doi.org/10.5281/zenodo.10447666}. \DOIprefix\doi{10.5281/zenodo.10447666}.

\end{thebibliography}
